\documentclass[aps,reprint, prl, superscriptaddress, showpacs,floatfix]{revtex4-1}

\usepackage{graphicx}
\usepackage{epstopdf}
\usepackage{amsmath}
\usepackage{amssymb}

\newcommand{\ppin}{P_{\mathrm{pin}}}
\newcommand{\ploose}{P_{\mathrm{loose}}}

\begin{document}
\title{Disorder strength and field-driven ground state domain formation in artificial spin ice: experiment, simulation and theory}

\author{Zoe~Budrikis}
\email{zoe.budrikis@gmail.com}
\affiliation{School of Physics, The University of Western Australia, 35 Stirling Hwy, Crawley 6009, Australia}
\affiliation{Istituto dei Sistemi Complessi CNR, Via Madonna del Piano 10, 50019 Sesto Fiorentino, Italy}
\affiliation{SUPA School of Physics and Astronomy, University of Glasgow, Glasgow G12 8QQ, United Kingdom}
\author{J.P.~Morgan}
\affiliation{School of Physics $\&$ Astronomy, University of Leeds, Leeds LS2 9JT, United Kingdom}
\author{J.~Akerman}
\affiliation{School of Physics $\&$ Astronomy, University of Leeds, Leeds LS2 9JT, United Kingdom}
\affiliation{Instituto de Sistemas Optoelectr\'{o}nicos y Microtecnolog\'{i}a (ISOM), Universidad Polit\'{e}cnica de Madrid, Ciudad Universitaria s/n, Madrid 28040, Spain}
\author{A.~Stein}
\affiliation{Center for Functional Nanomaterials, Brookhaven National Laboratory, Upton, New York 11973, USA}
\author{Paolo~Politi}
\affiliation{Istituto dei Sistemi Complessi CNR, Via Madonna del Piano 10, 50019 Sesto Fiorentino, Italy}
\affiliation{INFN Sezione di Firenze, Via G Sansone 1, 50019 Sesto Fiorentino, Italy}
\author{S.~Langridge}
\affiliation{ISIS, Rutherford Appleton Laboratory, Chilton, OX11 0QX, United Kingdom}
\author{C.H.~Marrows}
\email{C.H.Marrows@leeds.ac.uk}
\affiliation{School of Physics $\&$ Astronomy, University of Leeds, Leeds LS2 9JT, United Kingdom}
\author{R.L.~Stamps}
\affiliation{SUPA School of Physics and Astronomy, University of Glasgow, Glasgow G12 8QQ, United Kingdom}

\begin{abstract}
Quenched disorder affects how non-equilibrium systems respond to driving. In the context of artificial spin ice, an athermal system comprised of geometrically frustrated classical Ising spins with a two-fold degenerate ground state, we give experimental and numerical evidence of how such disorder 
washes out edge effects, and provide an estimate of disorder strength in the experimental system.
We prove analytically that a sequence of applied fields with fixed amplitude is unable to drive the system to its ground state from a saturated state.
 These results should be relevant for other systems where disorder does not change the nature of the ground state.
\end{abstract}


\maketitle
Artificial spin ice~\cite{Wang:2006, Tanaka:2006, Qi:2008} consists of nanofabricated arrays of elongated magnetic dots that are small enough to be single-domain, but large enough to be athermal.
Each dot can be represented by a macro Ising spin, whose dynamics is governed by frustrated magnetostatic interactions and an external magnetic field.
Like other artificial systems~\cite{Davidovic1996, Kirtley2005, Han:2008}, artificial spin ice offers a setting in which to explore athermal, nonequilibrium dynamics, with the advantage that its microscopic configurations are easier to measure than those of, e.g, granular materials.

Understanding these dynamics is difficult as the study of nonequilibrium systems lacks a key ingredient: the probability
distribution of microstates, i.e. the Boltzmann factor,
which allows the determination of all relevant physical
quantities~\cite{Zwanzig2001}. Furthermore, athermal systems are intrinsically strongly out of equilibrium and the relevant distribution function is not obtainable from perturbations
of the equilibrium distribution. 
Artificial spin ice is an experimental example of such an athermal system,  with the additional ingredient of geometrical frustration. Furthermore, it has been shown~\cite{Moller:2006, Libal:2009, Kohli2011, Ladak:2010, Mengotti:2010, Daunheimer2011, Libal2011a, Pollard2012} that although temperature is not relevant, randomness does enter via quenched disorder, due to small unavoidable variations during fabrication. 

\begin{figure}
\begin{center}
\includegraphics[width=0.8\columnwidth]{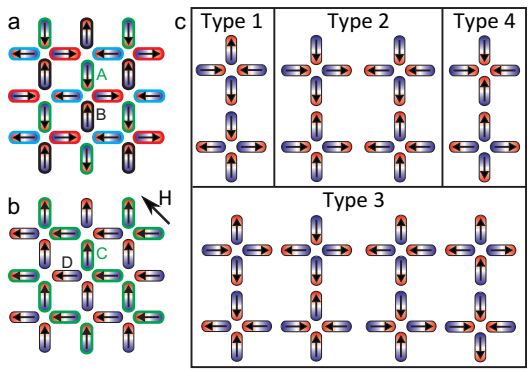}
\end{center}
\caption{
(a) A pure tiling of Type 1 vertices gives a ground state (GS) configuration. The other GS is obtained by flipping all spins.
(b) One of four diagonally polarized states (DPS) and the field direction used to obtain it.
(c) The 16 vertex configurations, grouped in order of increasing energy.}
\label{fig1}
\end{figure}

Previous studies have quantified disorder strength in artificial spin ices~\cite{Daunheimer2011, Mengotti:2010, Ladak:2010, Pollard2012}, but relatively little has been said about how it affects dynamics, especially in square ices. 
In particular, ideal square ices have a well-defined, two-fold degenerate ground state~\cite{Moller:2006, Mol:2009} (GS; see Fig.~\ref{fig1}(a)), but this has proven unattainable in experimental studies of field-driven demagnetization, which have yielded states with only short-range GS correlations~\cite{Wang:2006, Nisoli:2007, Ke:2008}. Simulation studies of a nanopatterned superconductor ``spin'' ice suggests that disorder is partially responsible~\cite{Libal:2009}, but its full role has not yet been elucidated. 

In this Letter, we examine the extent to which  disorder can disrupt ordering processes in artificial spin ice.
Our argument is a specific example of a broader problem of how disorder affects access to a set of degenerate states, and is also applicable to, e.g, antiferromagnetically coupled magnetic dots with a distribution of switching barriers~\cite{Cowburn2002, Imre2003, Thomson2006}. 
Our results complement previous studies of how disorder affects phase space~\cite{Budrikis2011networks, Budrikis2011njpnetworks}.

We present experimental studies in which rotating field protocols are used to drive dynamics in a square ice, in the manner simulated in Ref.~\onlinecite{Budrikis:2010}.
In those simulations, it was shown that -- in ideal systems -- a constant-amplitude rotating field can generate states with large domains of GS ordering via orderly invasion processes.
In our experiments, GS domains are smaller than predicted, which new simulations show is attributable to quenched disorder.
We prove analytically that disorder blocks GS access in real systems for \emph{any} field protocol with constant amplitude, by forcing multiple GS domains to nucleate.

An important related problem is that of how the basic ``step'' taken by the system as it moves through phase space affects the accessibility of states: for example, in Monte Carlo simulations of vertex models~\cite{Evertz1993}, single spin flip dynamics can ``freeze'' in regions of phase space far from the ground state, an issue not faced by loop-based dynamics. Indeed, this problem is general, appearing also in contexts such as domain wall creep dynamics~\cite{Rosso2001}. In this present Letter, we address the question of pathways to the GS in the context of  single spin flip dynamics, an approach motivated by the stochastic nature of spin flips in artificial spin ice, but leave open the possibility that dynamics governed by global moves may be different.

\begin{figure*}
\begin{center}
\includegraphics[width=2\columnwidth]{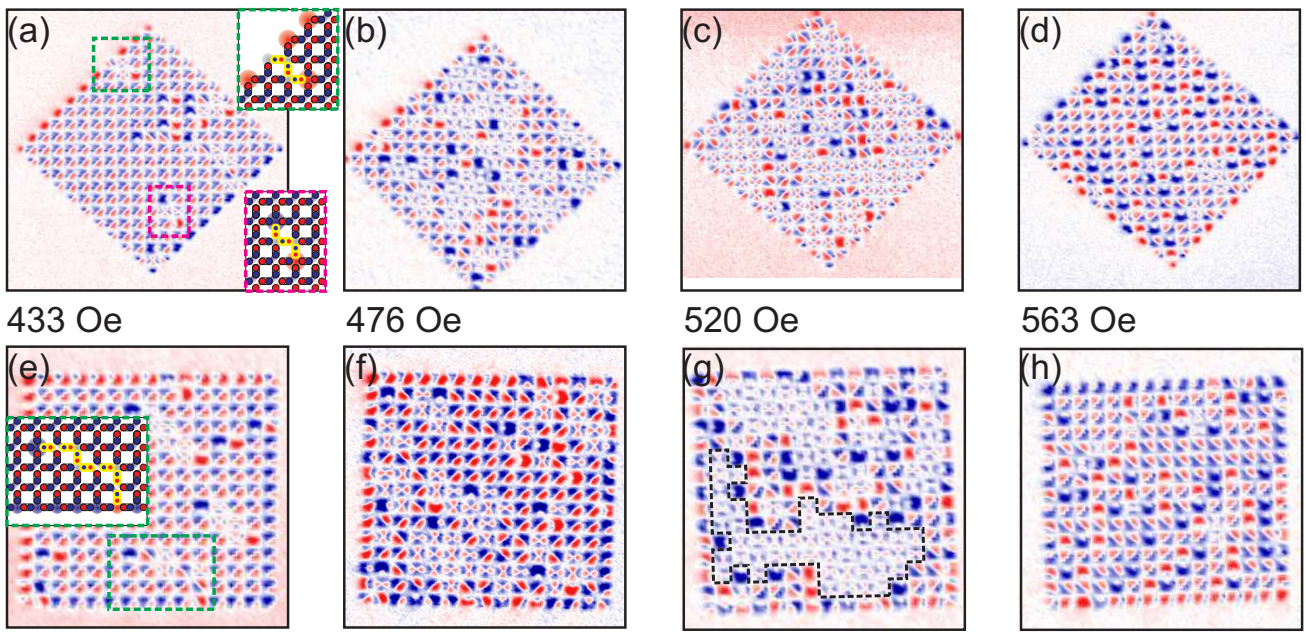}
\end{center}
\caption{MFM images of final states of open (a-d) and closed (e-h) arrays after rotation at selected hold fields. Insets indicate the corresponding island moments for selected areas of the array. A GS domain is outlined in (g). Images are false-colored using the software package WSxM~\cite{Horcas2007}.}
\label{fig2}
\end{figure*}

As in Ref.~\onlinecite{Budrikis:2010}, we study two array edge geometries. 
In ``open edge'' (``closed edge'') arrays, edge spins have an odd (even) number of nearest neighbors. These geometries are shown in the insets to Fig.~\ref{fig2}(a,e). In perfect systems, differences in coupling (of the order of nearest-neighbor interactions) at array edges cause edge geometry to ``select''  dynamics with distinct field dependence for the two array types. We will see that disorder disrupts this.

Five nominally identical arrays of each edge type were patterned on a single Si chip with electron beam lithography, as per Ref.~\onlinecite{Morgan:2011}. 
Islands were nominally 85 nm by 280 nm on a lattice of 400 nm constant, with a thin film structure of Cr(2nm)/Permalloy(30nm)/Al(2nm), forming moments of $\sim10^{6}\mu_B$, with nearest neighbor coupling of $\sim10$ Oe. 
As is common in demagnetization studies~\cite{Ke:2008, Phatak2011}, a large in-plane field $H = 2$ kOe was applied along a diagonal symmetry axis to prepare a diagonally polarized state (DPS; see Fig.~\ref{fig1}(b)). The field was then reduced to a hold value $H_{\rm h}$ and the sample was rotated in-plane. We studied hold fields at 22 Oe increments between 411 Oe and 606 Oe.
After hundreds of rotations -- more than enough to reach a predicted steady state \cite{Budrikis:2010} -- the field was ramped to zero at a rate of $\sim$ 10,000 Oe/s (compared to a rotation period of $\sim 30$ ms). 
We have confirmed by simulation that this ramp-down does not cause the demagnetization effects seen for slow-ramp protocols~\cite{Wang:2007, Moller:2006}, because the field range over which non-trivial dynamics can occur, $H_{\max}-H_{\min}\approx150$ Oe, is crossed within a single rotation. 
For large $H_{\rm h}$, the field angle at which ramp-down starts influences the final configuration, as described below, but its exact value is unimportant in general.
For each $H_{\rm h}$, remanent states were imaged by MFM, which shows a pole at each island end, indicated by the red/blue contrast in Fig.~\ref{fig2}, confirming that the islands are single-domain.

Array configurations can be conveniently represented in terms of vertices. Figure~\ref{fig1}(c) shows the sixteen possible vertex configurations grouped into types based on energy \cite{Wang:2006}. The GS is a chess-board tiling of Type 1 vertices, and the DPS that we use as an initial configuration is a uniform tiling of one of the Type 2 vertices. Dynamics on DPS or GS backgrounds can be described in terms of the motion of Type 3 vertices \cite{Mol:2009, Budrikis:2010, Mol:2010, Morgan2011a}. The population of Type 1 vertices serves as an indicator of the level of GS ordering \cite{Libal:2009, Budrikis:2010}.

\begin{figure}
\begin{center}
\includegraphics[width=\columnwidth]{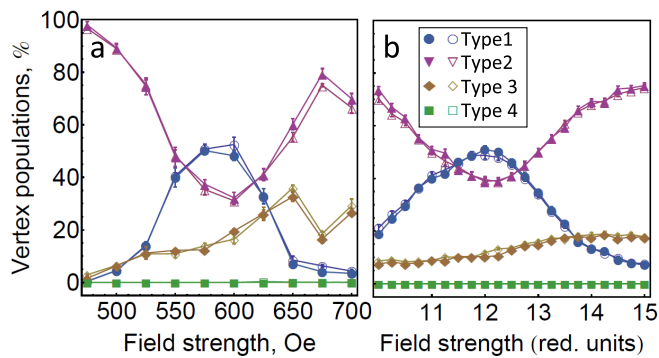}
\end{center}
\caption{Vertex populations \textit{vs} hold field for (a) experiment and (b) theory. 
Symbols represent vertex types as shown in the legend, with open (closed) symbols for open (closed) edge arrays. Each data point is the average over several runs; error bars represent the standard error. 
}
\label{fig3}
\end{figure}

Figures~\ref{fig2} (a-d) and (e-h) show example MFM images from the $H_{\rm h}$ series, for open and closed edge arrays respectively. Key configurations are mapped schematically (insets) in terms of dipole moments and vertex type. Note that images for different $H_{\rm h}$ are not all taken from the same array.
Average fractional vertex populations for each $H_{\rm h}$ are tracked in Fig.~\ref{fig3}(a).
We make two general observations. First, the population statistics and net magnetizations (not shown) are almost identical for open and closed arrays for all $H_{\rm h}$, indicating that edge effects are suppressed, which we show below is caused by disorder. Second, the maximum Type 1 population, found at $H_{\rm h} = 520$ Oe, is significantly reduced compared to the predictions of the ideal model \cite{Budrikis:2010}. 

Examining the configurations attained, we see that for $H_{\rm h} < H_{\rm min} \approx 410~\rm{Oe}$, the field does not affect the initial DPS. For $410~{\rm Oe} \le H_{\rm h} \le 455~{\rm Oe}$ the configurations at remanence consist of chains of reversed moments on a background of the initial DPS, as seen in Fig.~\ref{fig2}(a,e). It is clear from the MFM that most chains are nucleated in the bulk, presumably at sites with low switching barrier~\cite{Libal:2009}. 
The chains are similar to those reported previously in dc field experiments \cite{Ladak:2010,Mengotti:2010,Morgan2011a, Daunheimer2011, Pollard2012}. In those experiments, they occur via bulk nucleation, cascading and pinning under the influence of interactions and disorder.

As $H_{\rm h}$ is increased to 476 Oe, small GS domains form, and mixed GS/DPS phases are found, as seen in Fig.~\ref{fig2}(b,f). Near 520 Oe, increasing numbers of moments are reversed from the initial DPS, the magnetization approaches 0, and all four Type 2 vertices reach similar populations -- the memory of the initial DPS is lost. The GS domain size increases, and the Type 1 populations reach a maximum of 50\%, as seen in Fig.~\ref{fig2}(c,g) and Fig.~\ref{fig3}(a). An example GS domain is outlined in Fig.~\ref{fig2}(g). Domain wall structures separating GS domains, similar to those caused by thermal ordering~\cite{Morgan:2011}, are also observed. 

Increasing $H_{\rm h}$ further rapidly suppresses GS order, as Zeeman energy dominates and DPS ordering that couples to the field is preferred. Only moments with large switching barriers can pin; the rest align with the field. This is evident in the increasingly polarized states observed in Fig.~\ref{fig2}(d,h). The magnetization  direction is determined by the field direction at which ramp-down occurred. We estimate $H_{\rm max}$ to be 560 Oe; above this field, no significant GS domains are found, and simulations indicate that any GS ordering is picked up during ramp-down.

We now use numerical simulations to establish that the above observations can be explained by quenched disorder, and to estimate its strength relative to other energies in the system. 
In our simulations, the Ising spin $i$ flips if the total field acting on it, comprising the external field and dipolar interactions with all other islands, exceeds the threshold 
\begin{equation}
\label{switching}
\vec h_{\mathrm{tot}}^{(i)}\cdot \hat m_i < -h_c^{(i)},
\end{equation}
where $\hat{m}_i$ is a dimensionless unit vector along the spin direction and $h_c^{(i)}$ is the island's switching barrier, which, in a perfect system, is the same for all islands. This threshold-based model~\cite{Moller:2006, Remhof:2008, Mengotti:2010, Ladak:2010} has a $\cos\theta$ angular dependence, 
and is appropriate for describing the Zeeman-energy-driven propagation of domain walls, such as occurs during reversal of dots with dimensions similar to ours~\cite{Wernsdorfer2006}, in which domain wall nucleation is assisted by the curling of magnetization at island ends~\cite{Phatak2011}. 
(For smaller dots, Stoner-Wohlfarth switching~\cite{Stoner1948, Thirion2003} would be more realistic.)
Like other authors~\cite{Ladak:2010, Mengotti:2010, Daunheimer2011, Pollard2012}, we implement disorder by taking the $h_c^{(i)}$ from a Gaussian distribution with standard deviation $\sigma$; we show elsewhere that this type of disorder behaves similarly to disorder in interactions~\cite{Budrikis2011disorder}. 
 We work in reduced units where the nearest-neighbor dipole coupling is $1.5$ relative to $M^2/4\pi\mu_0$ ($M$ is the island net moment) and the mean $h_c$ value is $11.25$. 

We simulate a protocol in which the field rotates with constant amplitude $h$ and angular step $d\theta=0.01$ radians for 10 cycles, long enough to obtain a steady state. 
In line with experiments, the field is then ramped down over half a cycle to $h=8$, a field strength too low to induce dynamics.
At each field application, the system evolves by flipping single spins according to criterion~\eqref{switching} until no further flips are possible. 

We find good agreement with experimental vertex populations -- in terms of general trends, peak $n_1$ value, and lack of dependence on edge geometry -- when disorder is in the ``strong disorder regime'' of  Ref.~\onlinecite{Budrikis2011disorder}. For example, Fig.~\ref{fig3}(b) shows results for $\sigma=1.875$, a distribution width equal to $125\%$ the nearest neighbor coupling, and large enough to suppress edge effects. 
This value is in agreement with the value of $\sigma=60$ Oe, relative to a mean switching field of $320$ Oe, given by Pollard \textit{et al}~\cite{Pollard2012}, who studied arrays similar to ours. 

Disorder in our simulations is an \emph{effective} switching dispersion that incorporates effects from disorder in switching characteristics and interactions. As a point of comparison, if there was no disorder in interactions and island critical fields were directly proportional to their volume, $\sigma=1.875$ would correspond to a standard deviation in island linear dimensions of $5\%$.  Alternatively, if disorder originated only from fluctuations in nearest-neighbor interactions, the disorder would correspond to a standard deviation of $40\%$~\cite{Budrikis2011disorder}. 
We have been able to measure the standard deviation in island dimension: the value $\sim 1\%$ indicates that both types of disorder are present.

Having seen that a rotating field does not drive a disordered system from the DPS to a GS, we now ask: to what extent is this inherent to the nonequilibrium driven dynamics of a frustrated system, and what is the role played by disorder? 
Here we prove that when disorder is present no protocol with constant field amplitude can force a single GS ordering to cover the array, if disorder is strong enough or the system is large enough. 

A key ingredient of our argument is the two-fold degeneracy of the GS, which allows for separate GS domains to form. Thus, we require that unlike in, e.g, the random field Ising model (in which the GS becomes \emph{more} accessible for stronger disorder~\cite{Alava2005}), disorder should not change the nature of the GS. This is true for switching field disorder.

We consider two mechanisms by which formation of separate GS domains can occur. We calculate an upper bound on the probability $P(\text{not blocked})$ that neither mechanism operates; $P(\text{blocked})=1-P(\text{not blocked})$ is a lower bound on the probability the GS is blocked. We outline the argument here, and give details as Supplemental Material \cite{SM}.

The first mechanism depends on the initial state being a DPS. To drive a DPS to a GS, half the spins must be flipped, and the spins of the DPS can be divided into two groups based on their alignment with either GS.  Suppose one spin from each of the two groups is pinned and remains always in its initial state,  e.g, spin C and D in Fig.~\ref{fig1}(b) . A single GS cannot contain both C and D in their initial states, but a configuration with two GS domains can.

The GS is \emph{not} blocked by pinning only if at least one of the two groups contains no pinned spins. Then, the first terms in the expression for $P(\text{not blocked})$ depend on the probability, $\ppin$, of a spin being pinned -- that is, the probability the spin's switching barrier is so high it cannot flip, even in a maximally unfavourable local environment. This depends on field strength $h$. Two limits are $\ppin\to1$ as $h\to0$ and $\ppin\to0$ as $h\to\infty$.

The second mechanism of GS blocking does not rely on an initial DPS. If two spins that are antiparallel in the GS -- e.g, spin A and B in Fig.~\ref{fig1}(a) -- are ``loose'' and always align with the external field, then the GS is blocked. This gives a second set of terms in the expression for $P(\text{not blocked})$. $\ploose$, the probability a spin aligns with an external field even when its neighbors are GS ordered, has the limits $\ploose\to1$ as $h\to\infty$ and $\ploose\to0$ and $h\to0$.

\begin{figure}
  \centering
  \includegraphics[width=0.8\columnwidth]{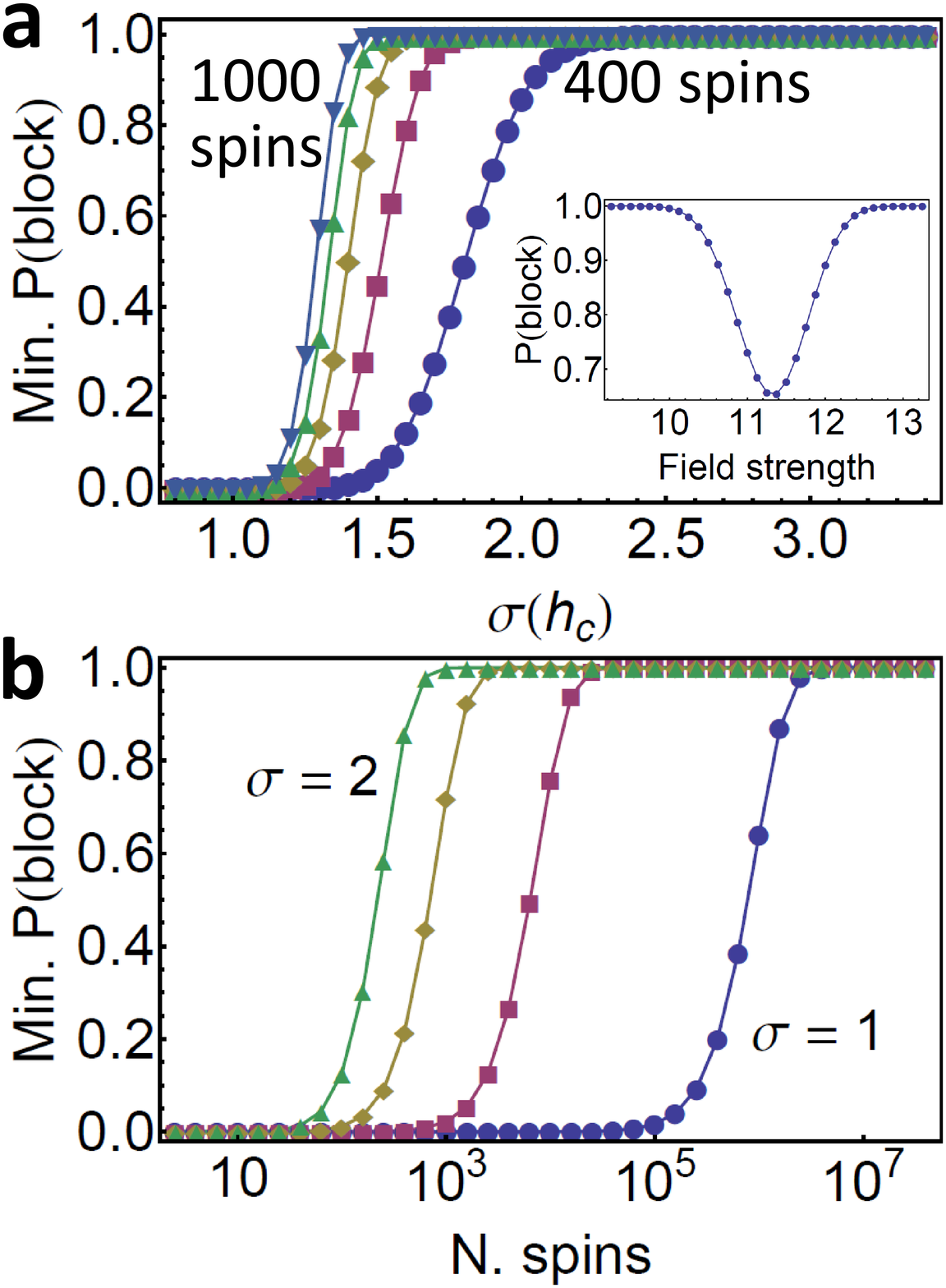}
  \caption{
  The minimum GS blocking probability \textit{vs} (a) disorder strength for arrays containing 400--1000 spins; and (b) array size for disorder strengths from $1.0$ to $2.0$.
  Inset: $P(\text{blocked})$ \textit{vs} field strength, for a $20\times20$ array with a Gaussian distribution of $h_c$ with standard deviation $1.875$.
  Numerical values are in the same reduced units used in the rest of this Letter.
  }
  \label{Pblock}
\end{figure}

The probability that the GS is blocked is
\begin{equation}
\label{blocking}
\begin{split}
P(\text{blocked})=1-\bigl[2 (1-\ppin)^{n/2}-(1-\ppin)^n\bigr]\\
	\times \bigl[4 (1-\ploose)^{n/2}(1-(1-\ploose)^{n/4})^2 \\
	+ 4 (1-\ploose)^{3n/4}(1-(1-\ploose)^{n/4})+(1-\ploose)^n \bigr].
\end{split}
\end{equation} 
The inset to Fig.~\ref{Pblock}(a) shows $P(\text{blocked})$ \textit{vs} field strength, for the system studied in simulations. $P(\text{blocked})>65\%$ always. Figure~\ref{Pblock} shows that the minimum of $P(\text{blocked})$ grows rapidly with disorder strength and array size.  In the limit of an infinite system, finite probabilities of pinned and loose spins lead to finite populations of spins in both GS alignments, and the GS is necessarily blocked.

Because we have been conservative in our estimates of $\ppin$ and $\ploose$, these results are a lower bound on $P(\text{blocked})$. While large $P(\text{blocked})$ indicates the GS is inaccessible, small $P(\text{blocked})$ does not mean it can be reached: for example, ideal systems can jam~\cite{Budrikis:2010}.
Our results apply to any field protocol with fixed field amplitude, such as field protocols where the sense of rotation alternates. An open problem is whether protocols with varying field amplitude face similar blocking. 
Finally, we have shown that although rotating fields do not attain the GS, they do achieve a high level of GS ordering, pointing to questions about interplay between disorder and  optimization~\cite{Zarand:2002, Alava2005, Pal2006}.

\begin{acknowledgments}
We thank Shawn Pollard and Yimei Zhu for useful discussions about disorder strength.
Funding was provided by the Australian Research Council and the Worldwide University Network (Z.B. and R.L.S.), INFN and the Hackett Foundation (Z.B.), and EPSRC and the Centre for Materials Physics and Chemistry at STFC (J.P.M. and C.H.M.).
Research carried out in part at the Center for Functional Nanomaterials, Brookhaven National Laboratory, which is supported by the U.S. Department of Energy, Office of Basic Energy Sciences, under Contract No. DE-AC02-98CH10886.
\end{acknowledgments}

% \bibliography{c:/Users/Zoe/Documents/library}


\end{document}